\documentclass[twocolumn]{aastex63}

\newcommand\om{`Oumuamua}

\newcommand\htwo{H\textsubscript{2}}

\usepackage{comment}

\shorttitle{Solid Hydrogen Objects from Starless Cores}
\shortauthors{Levine and Laughlin}

\graphicspath{{./}{figures/}}

\begin{document}

\title{Assessing the Formation of Solid Hydrogen Objects in Starless Molecular Cloud Cores}

\correspondingauthor{W. Garrett Levine}
\email{garrett.levine@yale.edu}

\author[0000-0002-1422-4430]{W. Garrett Levine}
\affiliation{Yale University \\
52 Hillhouse, New Haven, CT 06511, USA}

\author[0000-0002-3253-2621]{Gregory Laughlin}
\affiliation{Yale University \\
52 Hillhouse, New Haven, CT 06511, USA}

\begin{abstract}

The properties of the first-discovered interstellar object (ISO), 1I/2017 (\om), differ from both Solar System asteroids and comets, casting doubt on a protoplanetary disk origin. In this study, we investigate the possibility that it formed with a substantial \htwo{} ice component in the starless core of a giant molecular cloud. While interstellar solid hydrogen has yet to be detected, this constituent would explain a number of the ISO's properties. We consider the relevant processes required to build decameter-sized, solid hydrogen bodies and assess the plausibility of growth in various size regimes. Via an energy balance argument, we find the most severe barrier to formation is the extremely low temperature required for the favorability of molecular hydrogen ice. However, if deposition occurs, we find that the turbulence within starless cores is conducive for growth into kilometer-sized bodies on sufficiently short timescales. Then, we analyze mass loss in the interstellar medium and determine the necessary size for a hydrogen object to survive a journey to the Solar System as a function of ISO age. Finally, we discuss the implications if the \htwo{} explanation is correct, and we assess the future prospects of ISO science. If hydrogen ice ISOs do exist, our hypothesized formation pathway would require a small population of porous, $100\,\mu\text{m}$ dust in a starless core region that has cooled to $2.8\,\text{K}$ via adiabatic expansion of the surrounding gas and excellent shielding from electromagnetic radiation and cosmic rays.

\end{abstract}

\section{Introduction} \label{sec:intro}

The observed properties of the interstellar object (ISO) 1I/2017 U1 (\om) do not fit neatly within the known astronomical taxonomy. In the Solar System, minor bodies are usually classified as asteroids or comets, with the latter often experiencing force from solar-driven sublimation of volatiles. Optical and IR images of \om{} found no coma \citep{meech2017brief, jewitt2017observation, trilling2018spitzer}, yet analysis of its trajectory revealed non-gravitational acceleration \citep{micheli2018acceleration}. Additionally, dramatic light-curve variations implied that its shape was approximately that of a 6:6:1 oblate ellipsoid \citep{mashchenko2019modelling}. Together, these attributes suggest that \om{} is a novel type of object and distinct from both Solar System comets as well as the extrasolar comet 2I/Borisov \citep{jewitt2019borisov}.

To reconcile \om's trajectory with a non-volatile composition, some studies have suggested radiation pressure as the non-gravitational force. For example, \cite{bialy2018radiation} propose an unusually-thin minor axis, whereas others investigate ultra-porous objects with bulk density $\rho \sim 10^{-5}\,\mathrm{g/cm}^{3}$. Specifically, \cite{moro2019fractal} considers coagulation of icy grains in a protoplanetary disk, and \cite{luu2020oumuamua} examines diffusion-limited aggregation in an exo-comet's tail.

Radiation pressure explanations generally assume implicitly that \om's origin was an exoplanetary system: a ``disk product" hypothesis. However, this birthplace has difficulties with explaining either \om's kinematics or the inferred number density as representative of a Galactic reservoir of ISOs. First, an inbound velocity near the Local Standard of Rest (LSR) is consistent with an object age of less than $100\,\text{Myr}$ \citep{mamajek2017kinematics, hallatt2020dynamics}. If disk products form consistently throughout Galactic history, \om{} would have coincidentally been selected from the population's youngest 1\%. Furthermore, Pan-STARRS' detection of any ISO within 0.2 AU of Earth implies number densities such that the average star must eject at least 20 M$_{\earth}$ of material in disk-product hypotheses \citep{do2018interstellar}. This large productivity is in marginal tension with planetary formation models.

As an alternative to radiation pressure, \cite{seligman2020h2} propose that outgassing of molecular hydrogen acquired in a giant molecular cloud (GMC) could account for \om's non-gravitational acceleration and non-detection of common volatiles. Additionally, it explains the inbound kinematics and extreme aspect ratio as consequences of erosion during the finite lifetime of solid \htwo{} in the interstellar medium (ISM) \citep{domokos2017explaining}. Because GMCs are a larger mass source than exoplanetary systems, number density constraints are easily satisfied. Finally, a ``cloud product" would corroborate \om's strong dynamical correlation with the surrounding Columba and Carina stellar associations \citep{gaidos2018binary, hallatt2020dynamics}, the latter of which may be as young as $25\,\text{Myr}$ as determined by recent spectroscopic evidence \citep{schneider2019acronym}. Furthermore, the cloud product hypothesis implies an origin for the ISO that was independent of a particular host star.

Because \htwo{} is the only proposed composition that can explain all of \om's properties as generic to a population of objects, it is imperative that potential formation pathways be closely-examined. This study evaluates the plausibility of \htwo{} ice ISOs originating in GMCs, a possibility raised soon after \om{}'s detection by \citet{fuglistaler2018solid}. Specifically, we focus on regions that are unproductive in star-formation since these will host the coldest temperatures. These comparatively-quiescent GMCs, in contrast to the solar birthplace, host numerous starless cores \citep{vazquez2005lifetimes}.

Solid hydrogen has previously been examined as a possible source of dark matter \citep{pfenniger1994darkII, white1996dark}. However, current $\Lambda$CDM cosmology requires a substantial non-baryonic component to the Universe's matter density in addition to any remaining unidentified hydrogen \citep{bergstrom2000dark}. Assuming ISO number density consistent with \cite{do2018interstellar}, a population of \om-like objects would not be dynamically important for the Galaxy. Thus, the channel of \htwo{} production that we investigate would have no immediate ramifications for the topic of dark matter.

Theoretically, \cite{fuglistaler2018solid} analyze gravitationally-unstable fluids near solid-vapor equilibrium, arguing that starless cores may spawn objects ranging from small condensates via deposition to multi-kilometer ISOs via coagulation. They find scaling laws at high pressure using molecular dynamics simulation, then extrapolate to assess GMCs. Similarly, \cite{walker2019cosmic} consider self-gravitating interstellar gas clouds, showing that heat transport via convection may be conducive to lowering ambient temperatures beneath the \htwo{} solid-vapor equilibrium and producing objects with a variety of characteristic sizes. In this study, we focus on the feasibility of producing \om-like objects in starless cores.

Observationally, \cite{sandford1993spectroscopic} attributed a detected $2.415\,\mu\mathrm{m}$ feature in the direction of $\rho$-Oph to solid \htwo{} embedded in interstellar ices. However, the result is unconfirmed and has recently been noted as an upper bound on concentration \citep{boogert2015observations}. Our study here, however, considers bulk object formation as opposed to processed ices.

We divide hypothetical cloud product formation into distinct size regimes, then consider the processes affecting growth in each one. In Section \ref{section:nucleation}, we consider the starless core energy balance and the prospects for solid hydrogen deposition on dust. Then, Section \ref{section:deposition} constrains the rates of \htwo{} and dust accretion onto an isolated object. Next, Section \ref{section:gravitysnowball} analyzes systems with multiple aggregating ``hailstones," which may form planetesimal-sized, snowball-like ISOs. Section \ref{section:ism} describes erosion in the ISM, which determines the object's metallicity upon encountering the Solar System. Finally, Sections \ref{sec:discussion} and \ref{section:conclusion} address the ramifications of \om{} as a cloud product. Because of the stringent requirements for deposition, confirmation of \htwo{} in future ISOs would probe previously-undetected interstellar conditions.

\section{Solid Hydrogen Deposition}
\label{section:nucleation}

\subsection{Thermodynamic Requirements for Solid Hydrogen}

\begin{figure}
\plotone{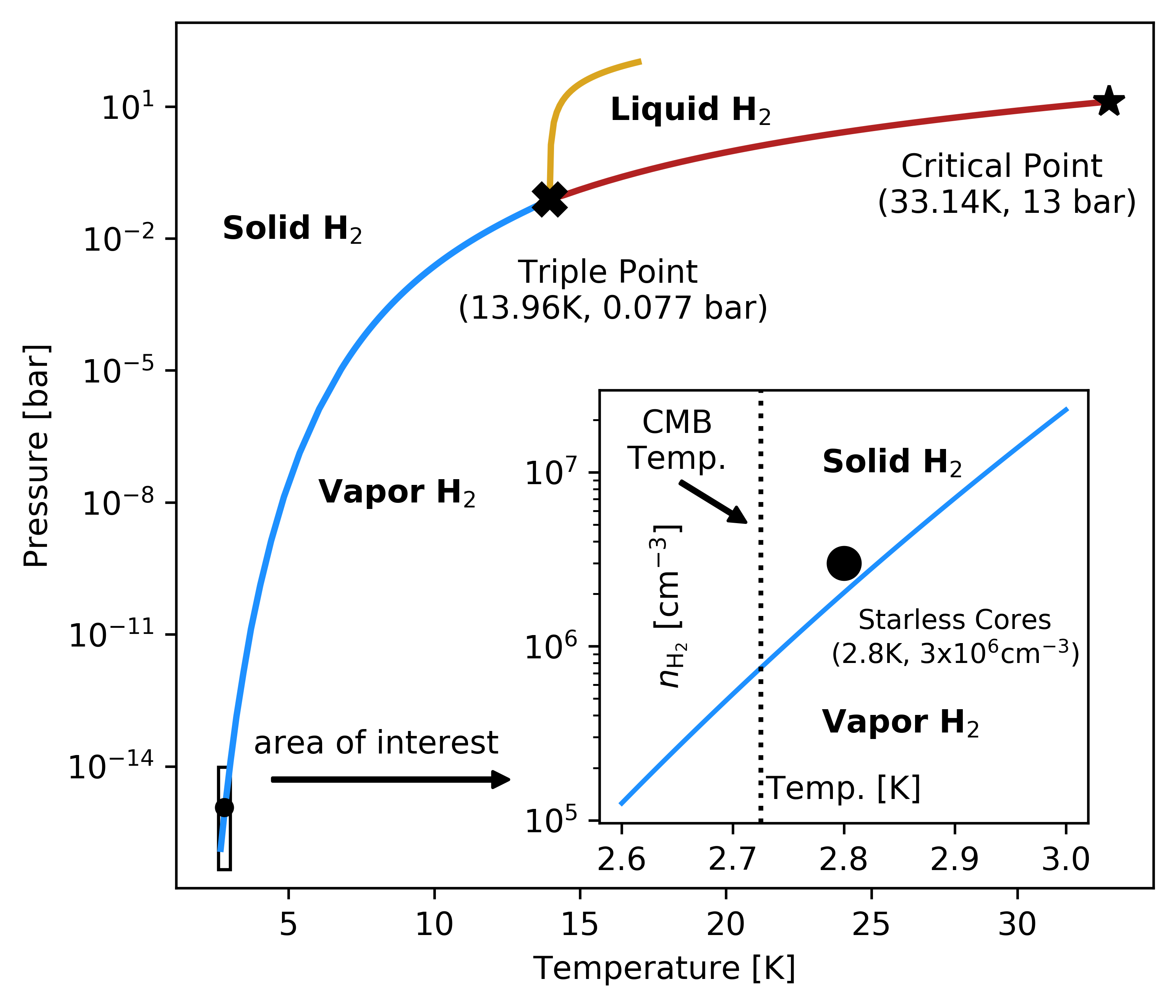}
\caption{Phase diagram for H\textsubscript{2} \citep{airliquide1976gas, anderson1989physicist}, with the solid circular mark corresponding to the fiducial starless core from Table \ref{tab:conditions}.
\label{fig:phasediagram}}
\end{figure}

Figure \ref{fig:phasediagram} shows the phase diagram for pure \htwo{}. If \om-like objects contain molecular hydrogen, the outgassing material could only be acquired where the solid phase is favored, motivating our scrutiny of the coldest, densest interstellar environments: starless cores in GMCs. Although the presence of He may modify the equilibrium lines, we assume the effect is negligible because of helium's inert nature. In support of this premise, \cite{fuglistaler2016formation} consider deposition in two-component fluids and find qualitatively similar phenomena compared to a pure system.

Nonetheless, the approximately exponential temperature dependence of solid-vapor equilibrium \citep{walker2013snowflake} means that slight deviations from Figure \ref{fig:phasediagram} may critically affect the feasibility of deposition. While previous studies of \htwo-He mixtures have often considered conditions analogous to giant planet interiors \citep{militzer2005hydrogen}, laboratory data for ISM-like temperature and pressure will be necessary if hydrogen cloud products are found in the future.

\subsection{Extreme Conditions in Starless Cores}

\begin{deluxetable*}{cccc}
\label{tab:conditions}
\tablecaption{Summary of assumed conditions in an extremely cold, dark, dense starless core.}
\tablewidth{0pt}
\tablehead{
\colhead{Symbol} & \colhead{Quantity} & \colhead{Value} & \colhead{Source}
}
\startdata
$T_{\mathrm{sc}}$ & Gas Temperature & 2.8 K & \cite{kong2016deuterium}\\
$n_{\mathrm{\htwo}}$ & Number Density of \htwo & $3.0 \times 10^{6}$ cm\textsuperscript{-3} & \cite{ivlev2015interstellar} \\
$\phi_{\mathrm{d}}$ & Dust/Gas Mass Ratio & $1.99 \times 10^{-2}$ & \cite{draine2011book} \\
$f_{\mathrm{He}}$ & Helium Number Fraction & 0.2 & \cite{draine2011book} \\
$A_{\text{v}}$ & Visual Extinction & 200 mag & \cite{draine2011book}\\
$R_{\mathrm{sc}}$ & Starless Core Radius & $5.0\times 10^{3} \text{ AU}$ &
\cite{weidenschilling1994coagulation} \\
$c_{\mathrm{s}}$ & Sound Speed & $1.0 \times 10^{4} \text{ cm/s}$ & \cite{weidenschilling1994coagulation}\\
$\tau_{\mathrm{sc}}$ & Starless Core Lifetime & $3 \times 10\textsuperscript{6}$ yr & \cite{kong2016deuterium}\\
$\tau_{\mathrm{cl}}$ & GMC Lifetime & $3 \times 10\textsuperscript{7}$ yr &
 \cite{kong2016deuterium}\\
\enddata
\end{deluxetable*}

GMC morphology is often governed by the turbulence that is ubiquitous in the ISM \citep{vazquez2005lifetimes}, so density and pressure within a given cloud can vary by orders-of-magnitude. Typical $30\,\mathrm{Myr}$ GMC lifetimes are too short for conditions to reach thermodynamic equilibrium \citep{shu1987star}, and the gas is often described as ``clumpy" \citep{draine2011book}. Because starless cores are often colder and better-shielded from electromagnetic radiation than the surrounding cloud, they are the optimal locations to investigate solid hydrogen deposition \citep{fuglistaler2018solid}.

Table \ref{tab:conditions} details our fiducial assumed conditions, and the corresponding point is marked in Figure \ref{fig:phasediagram}. Nominally, these environments would be unstable to the canonical Jeans' gravitational collapse, but we assume that magnetic field pressure supports the core for a few $10^{6}\,\text{yr}$ before re-expansion \citep{vazquez2005lifetimes, kong2016deuterium}. We equate the dust/gas mass ratio to the cloud metallicity, which implies that all elements other than H and He are in the solid phase. While much ISM hydrogen is atomic, catalysis on dust grains leads to the molecular form dominating in cores \citep{hollenbach1971surface}.

Importantly, our assumed temperature favors the solid phase of molecular hydrogen, as indicated by the phase diagram in Figure \ref{fig:phasediagram}. Sustained $T_{\text{sc}}$ this close to that of the cosmic microwave background (CMB) has yet to be seen, but excitation temperatures around $3.5\,\text{K}$ in M31 and $4.0\,\text{K}$ in our Galaxy have been reported \citep{loinard1998cold, kong2016deuterium}. GMC observations continue to find density structure at the smallest resolvable scales; thus, the coldest, densest regions have yet to be constrained \citep{vazquez2005lifetimes}. Because observations have yet to detect temperatures below the solid-vapor equilibrium line in starless cores, we must instead examine the most frigid conditions that these regions could theoretically support.

\subsection{Dust Grain Energy Balance in Starless Cores}

Even if thermodynamics supports the formation and persistence of solid \htwo, there must be surfaces on which deposition begins. Dust is the most viable candidate, but its temperature $T_{\mathrm{gr}}$ will not necessarily equilibrate to that of the surrounding gas. In starless cores, grains experience the following energy sources \citep{draine2011book}:

\begin{enumerate}
    \item Any power from stars in the Galaxy -- the interstellar radiation field (ISRF) -- at wavelengths for which the surrounding GMC is optically-thin.
    \item Blackbody CMB radiation with $T_{\mathrm{cmb}} = 2.725\:\text{K}$.
    \item Infrared (IR) photons emitted by warm dust in the outer cloud layers.
    \item Inelastic gas-grain collisions working to equilibrate $T_{\mathrm{gr}}$ and $T_{\mathrm{sc}}$.
    \item Bombardment by relativistic cosmic rays (CRs).
    \item Heat of decay from short-lived radionuclides (SLRs) delivered from supernovae.
\end{enumerate}

Broadly, we classify these sources into incident radiation (items 1-3) and other mechanisms (items 4-6). We assume there are no field stars providing anisotropic flux, consistent with examining clouds with low star-formation rates. Furthermore, because the only heat sources for the gas are the ISRF, CRs, and collisions with dust, $T_{\text{gr}}$ drives the overall starless core energy balance at our assumed number density \citep{goldsmith2001molecular}. Therefore, constraining the dust temperature is imperative to assessing the formation of solid \htwo.

In general, these energy fluxes can vary with position in the Galaxy. Nonetheless, a young inferred ISO age and a low velocity versus the LSR indicate that \om{} has likely experienced interstellar conditions similar to those surrounding the Solar System today. Thus, the magnitudes of the incident power sources are better constrained than they would be for an old object.

\subsection{Photon Radiation Bath for Ideal Grains}

We first consider ideal grains characterized by the Mathis-Rumpl-Nordsieck (MRN) distribution, with $dn/da \propto a^{-3.5}$ describing dust size $a = 0.005-0.25\,\mu\text{m}$ \citep{mathis1977size}. Most of the mass resides in large grains, while nearly all of the surface area is in small grains \citep{draine2011book}. Furthermore, we consider spherical, non-porous particles with bulk density $\rho_{\text{d}} = 3\,\text{g/cm}^{3}$. While both silicate and graphite dust should reside in starless cores, the former will maintain a slightly lower temperature and be more permissive of deposition \citep{mathis1983interstellar}. With $\phi_{\text{d}}$ from Table \ref{tab:conditions}, we find number density $n_{\text{d}} = 1\times10^{-6}\,\text{cm}^{-3}$ for the largest grains of mass $m_{\text{d}} = 2\times10^{-13}\,\text{g}$.

At steady-state, the fluxes sum to zero: $\sum dE/dt = 0$. For an isolated grain in a radiation bath, we equate the absorbed $(dE_{\mathrm{a}}/dt)$ and emitted $(dE_{\mathrm{e}}/dt)$ power as

\begin{equation} \label{eq:ISRF}
    \frac{dE_{\mathrm{a}}}{dt} =
    f_{\mathrm{r}}A_{\mathrm{g}}\epsilon_{\mathrm{a}} = \sigma T_{\mathrm{gr}}^{4} A_{s}  \epsilon_{\mathrm{e}}
    = \frac{-dE_{\mathrm{e}}}{dt}\,,
\end{equation}

\noindent where $f_{\text{r}}$ is the incident radiation flux, $A_{\text{g}}$ and $A_{\text{s}}$ are the cross-section and surface areas of the particle, $\epsilon_{\text{a}}$ and $\epsilon_{\text{e}}$ are the absorption and emission efficiencies, and $\sigma$ is the Stefan-Boltzmann constant. We assume the ISRF at galactocentric distance $10\,\text{kpc}$ from \cite{mathis1983interstellar}, with $f_{\text{r}} = f_{\text{ISRF}} = 2.7\times10^{-2}\,\text{erg/cm}^{2}\text{/s}$ over all wavelengths. Without considering other power sources such as the CMB, a perfectly-efficient blackbody ($\epsilon_{\text{a}} = \epsilon_{\text{e}} = 1$) in only the Galactic ISRF would have temperature $T_{\text{gr}} = 3.3\,\text{K}$.

While this ISRF prohibits solid \htwo{} deposition, most incident radiation is eliminated in the outer cloud layers. \cite{hoang2020destruction} consider attenuation of all incident radiation by a single scaling factor, but we use wavelength-dependent extinction $A_{\lambda}$ from \cite{mathis1983interstellar}. Specifically, the quantities $\epsilon_{a}$ and $\epsilon_{e}$ are less than unity for MRN-sized grains and given by \cite{draine2011book} for silicate grains as

\begin{equation}\label{eq:efficiency}
    \epsilon \approx 1.4\times10^{-3}
    \Big[\frac{\lambda}{100\,\mu\mathrm{m}}\Big]^{-2}
    \Big[\frac{a}{0.1\,\mu\mathrm{m}}\Big]\,.
\end{equation}

\begin{figure}
\plotone{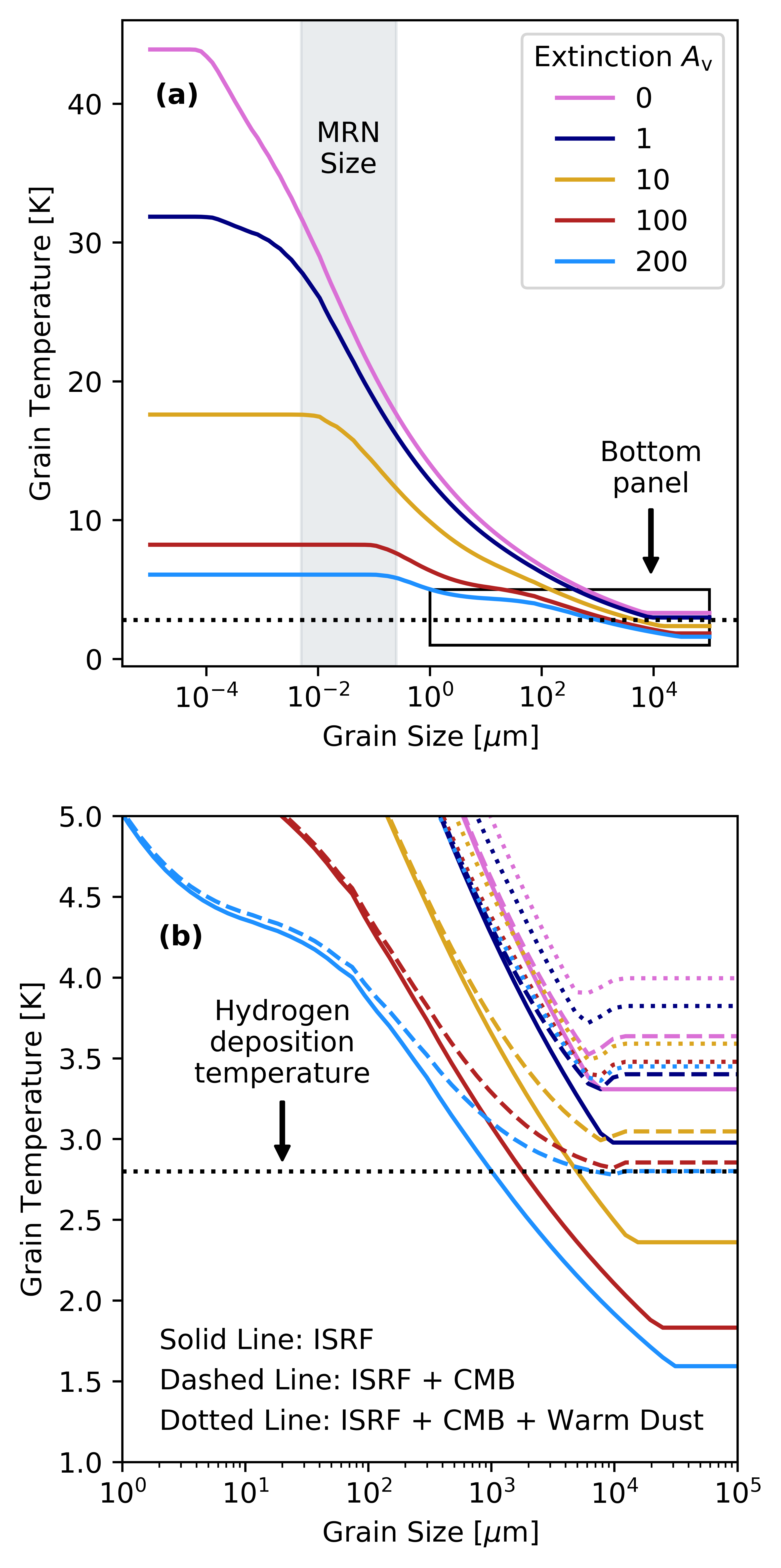}
\caption{Temperature of spherical ideal grains in a photon bath for various sources of incident radiation. \label{fig:graintemp}}
\end{figure}

\noindent Thus, the grain temperature from Equation \ref{eq:ISRF} is also size-dependent. With Equation \ref{eq:efficiency}, we can compute the average efficiency over the Planck function to find \citep{draine2011book}

\begin{equation} \label{eq:planckavg}
    \epsilon_{\mathrm{pl}} \approx 1.3\times10^{-6}
    \Big[\frac{a}{0.1\,\mu\mathrm{m}}\Big]
    \Big[\frac{T}{1\,\mathrm{K}}\Big]^{2}\,,
\end{equation}

\noindent which is $\epsilon_{\text{e}}$ for ideal particles. With these efficiencies, we plot the resulting $T_{\text{gr}}$ for varying grain sizes on the solid lines in Figure \ref{fig:graintemp}. MRN dust is too warm, so we display $a \gtrsim 1000\,\mu\text{m}$ where deposition may be feasible.

In addition, we must consider the CMB with $T_{\text{cmb}} = 2.725\,\text{K}$. Attenuation is negligible in starless cores at peak wavelength $\lambda_{\text{cmb}} \sim 1000\,\mu\text{m}$. We plot the modified $T_{\text{gr}}$ with the combined flux on the dashed lines of Figures \ref{fig:graintemp}. As expected, the CMB has a pronounced effect on the dust energy balance only as grain sizes approach $\lambda_{\text{cmb}}$.

Finally, we consider another source of photons: warm dust from the outer GMC layers. Grains located at lower $A_{\text{v}}$ are responsible for optical darkness in starless cores, and they re-emit energy primarily as IR. The spectrum and strength of this radiation depend on cloud geometry and other considerations. For example, \cite{mathis1983interstellar} superimposes two blackbody fields from graphite and silicate grains. Similarly, we use fields with $T_{1} = 16\,\text{K}$ and $T_{2} = 7\,\text{K}$ and dilution factors $W_{1} = 1\times10^{-3}$ and $W_{2} = 6\times10^{-3}$, with our deviations from \cite{mathis1983interstellar} attributed to galactocentric distance.

The effect of warm dust flux added to the ISRF and CMB on the energy balance of starless core grains is shown in the dotted lines of Figure \ref{fig:graintemp}. Deposition cannot occur on ideal grains in our fiducial starless core's radiation bath.

\subsection{Other Processes' Effects on Ideal Grains}

Next, we consider inelastic gas-grain collisions reducing the difference between $T_{\text{gr}}$ and $T_{\text{sc}}$ \citep{burke1983gas}. The effect is quantified by \cite{draine2011book} as

\begin{equation} \label{eq:collisionalheating}
    \frac{dE_{\mathrm{gas}}}{dt} = \sum_{i} n_{i}
    \Big[\frac{8kT_{\mathrm{sc}}}{\pi m_{i}}\Big]^{1/2}
    A_{\mathrm{g}}\alpha_{i} 2k\Delta T\,,
\end{equation}

\noindent where the sum is over all $i$ gaseous species with masses $m_{i}$, and $\Delta T = T_{\text{sc}} - T_{\text{gr}}$. Here, $\alpha_{i}$ quantifies the degree of inelasticity, the accommodation coefficient of \cite{burke1983gas}. For sticking fraction of the $i^{\text{th}}$ species $s_{i}$ and the rest of the collisions elastic, we find $\alpha_{i} = s_{i}$; otherwise, energy conservation requires that $\alpha_{i} > s_{i}$.

We consider the gas to contain only \htwo{} and He since other molecules freeze around $A_{\text{v}} \gtrsim 25$ \citep{mathis1983interstellar, goldsmith2001molecular}. With $\alpha_{i} = \alpha = 0.5$ \citep{draine2011book}, the \htwo/He ratio from Table \ref{tab:conditions}, and spherical grains, we simplify Equation \ref{eq:collisionalheating} to

\begin{equation} \label{eq:collisionssimple}
    \frac{dE_{\mathrm{gas}}}{dt} = 
    \Big[1.16 \frac{n_{\text{\htwo}}}{\sqrt{m_{\text{\htwo}}}}\Big]
    \Bigg[a^{2}\alpha \sqrt{32\pi k^{3}T_{\mathrm{sc}}}\Bigg]\Delta T\,.
\end{equation}

\noindent For MRN-sized dust, inelastic collisions do not effectively regulate $T_{\text{gr}}$ for $\Delta T < 10\,\text{K}$. Furthermore, the $a^{2}$ dependence of both Equations \ref{eq:ISRF} and \ref{eq:collisionssimple} in the ideal blackbody limit implies that gas-grain interactions are negligible for all particle sizes.

A violent means of kinetic energy transfer comes from cosmic ray (CR) bombardment, which heats the ambient gas and strikes the grains. \cite{white1996dark} considers heating of ISM dust and assumes $f_{\text{CR}} = 1.6\times 10^{-3}\,\text{erg/cm}^{2}\text{/s}$. Because the GMC attenuates some low-energy CRs, this estimate is an upper bound. Since $f_{\text{CR}}$ is smaller than the power delivered by the photon bath, it can be neglected in the energy balance.

Finally, we consider short-lived radionuclides (SLRs) synthesized in supernovae. For example, approximately $8\times10^{-6}$ of aluminum is Al-26, with half-life $7\times10^{5}\,\text{yr}$ and heat of decay $4\,\text{MeV}$ \citep{Gounelle2008isotopes}. Given cosmic Al abundance of $3\times10^{-6}$ \citep{draine2011book}, we find that the typical grain receives flux $1.5 \times 10^{-5}\,\text{erg/cm}^{2}\text{/s}$. Because clouds with low star-formation rates are not enhanced in radioactive material, we can safely ignore SLRs.

Both kinetic and radiogenic power are inconsequential, so the photon bath drives $T_{\text{gr}}$ in the ideal regime. Thus, the energy balance imposes a severe constraint for solid hydrogen deposition on MRN-like dust.

\subsection{Formation of Non-Ideal Grains}

Thus far, our underlying assumption has been ideal, sub-micron grains. While the MRN distribution fits interstellar extinction well, it does not preclude the existence of larger, porous, non-spherical dust. Instead of ``reddening" the ISRF, sub-millimeter particles are in the regime of geometric optics, extinguish all visual wavelengths equally, and appear spectroscopically ``grey," as understood via Equation \ref{eq:efficiency} \citep{draine2011book}.

Non-ideal particles might populate starless cores through either in-situ coagulation or delivery from an exogenous source. In particular, subsonic turbulence imparts relative velocities that usually do not shatter colliding dust aggregates, so dense cores naturally host grains larger than the MRN upper bound appropriate for other ISM dust \citep{ormel2009dust}. For example, numerical models from \cite{ormel2011dust} and \cite{weidenschilling1994coagulation} find that clouds may produce $100\,\mu\text{m}$ conglomerates within a few $10^{6}\,\text{yr}$.

Other studies have analyzed enhanced coagulation from electrostatic effects \citep{spitzer1941dynamics, ivlev2015interstellar}. Because the rates of positive charging via the photoelectric effect and negative charging via the ambient cold plasma both scale as $a^{2}$, the mean grain charge is independent of particle size and near zero for a range of plausible conditions. However, the random-walk nature of these charging pathways means that individual grains are charged; this condition is optimal for coagulation \citep{ivlev2015interstellar}. Oppositely-charged dust has increased collisional cross-sections from monopole effects at large separation due to net charge \citep{spitzer1941dynamics} and dipole effects at shorter distances due to induced charge distributions on conductive grains \citep{draine2011book}.

Another potential source of non-ideal particles could be delivery from the planetary nebulae of AGB stars. Specifically, \cite{sahai2017coldest} detected continuum millimeter emission in the Boomerang Nebula, which they attribute to sub-centimeter grains with total mass $5\times10^{-4}\,\text{M}_{\odot}$. Expulsion into the ISM may fracture particles, although conglomerate survival is unconstrained. Coincidentally, \cite{sahai2017coldest} also detect $1\,\text{K}$ excitation temperature in the adiabatically-expanding nebula, although the rapid outflow is unlikely to support deposition.

Characterization of a diffuse reservoir of coagulated interstellar particles is challenging, but \emph{Spitzer} has provided indirect evidence via GMC observations. A system of aggregates has smaller cross-sectional area versus free grains, so the dust extinction per hydrogen nucleon decreases with coagulation \citep{kim1996dust, ormel2011dust}. Additionally, limited direct measurements of presolar grains also have indicated that coagulated aggregates must exist. For example, \emph{Ulysses} and \textit{Stardust} both recovered porous, micron-sized aggregates \citep{landgraf2000ulysses, westphal2014evidence}.

While the assumption of idealized MRN grains provides a good model for diffuse interstellar space, starless cores may host conglomerates with more complicated geometries and compositions. Thus, we must assess the ability of these aggregates to support hydrogen deposition.

\subsection{Energy Balance of Non-Ideal Grains}

Non-ideal dust conglomerates, in addition to their large size, will be porous, anisotropic, and inhomogeneous. Because the material strength of individual small grains resists compaction in sub-millimeter aggregates, coagulated grains are characterized by fractal dimension $D_{\text{f}} \approx 2$ \citep{kataoka2013fluffy}. While the precise modifications to the energy balance are grain-specific, we can estimate the magnitude of the population-level deviations required for \htwo{} production.

For example, non-spherical geometry modifies the ratio $A_{\text{s}}/A_{\text{g}}$, thus altering $T_{\text{gr}}$. \cite{greenberg1971shape} consider cylindrical grains, showing that temperatures can decrease by 10\% versus the spherical limit. If this effect persists for irregularly-shaped dust, then $T_{\text{gr}}$ will be lower than in Figure \ref{fig:graintemp}. In addition, a porous, fractal configuration may increase the area radiating power while holding the cross-section intercepting flux close to constant.

Furthermore, conglomerates acquire impurities because dust does not necessarily coagulate with grains of similar composition. With all elements except H and He in the solid phase, the resulting imperfections may permit more vibrational modes for the grain than an idealized sphere. Furthermore, some of the impurity particles may efficiently emit IR radiation \citep{werner1969grain}. It must be noted, however, that grains which support \htwo{} ice due to inhomogeneities would have difficulty retaining these irregularities at macroscopic size.

Additionally, it's reasonable to assume that icy starless core grains may have non-zero albedos and reflect more incident radiation than their soot-like ISM counterparts. If particles hold sufficient solid hydrogen, the object albedo will be determined by that of \htwo{} ice.

Each of the aforementioned imperfections lowers $T_{\text{gr}}$. The largest coagulated grains ($a \gtrsim 100\,\mu\text{m}$) may be close to the deposition point even as spherical, non-porous blackbodies, so it's possible that some of the largest aggregates can support the phase transition. While deposition can be theoretically ruled-out in the ideal regime, the observational fit of a solid hydrogen component for \om{} warrants further investigation into the configuration of sub-millimeter, non-ideal dust in starless cores.

\subsection{Spacing of Deposition Sites within Starless Cores}

If hydrogen deposition does occur, the phase transition could only begin on non-ideal grain surfaces. Although the existence of sub-millimeter particles is unconstrained, we can place lower bounds on their typical separation. We assume that $a \gtrsim 100\,\mu\text{m}$ conglomerates permit deposition through a combination of the aforementioned non-ideal effects. While these aggregates do not reach the minimum temperatures from Figure \ref{fig:graintemp}, this size is a reasonable upper bound on coagulation \citep{weidenschilling1994coagulation}.

To begin, we express the number of dust grains with size $a_{\text{d}}$ needed to build a conglomerate with size $a_{\text{c}}$ and fractal dimension $D_{\text{f}}$ as \citep{moro2019fractal}

\begin{equation} \label{eq:conglomerate}
    N \approx \Big[\frac{a_{\mathrm{c}}}{a_{\mathrm{d}}}\Big]^{D_{\mathrm{f}}}\,.
\end{equation}

\noindent If fraction $\chi$ of the original MRN dust participates in coagulation and the resulting conglomerates are uniformly-spaced, then the average separation $l_{\text{c}}$ is

\begin{equation} \label{eq:spacing}
    l_{\mathrm{c}} \approx \Big[\frac{(N/\chi)}{n_{\text{d}}}\Big]^{1/3}
    \approx \Big[\frac{a_{\mathrm{c}}}{a_{\mathrm{d}}}\Big]^{D_{\mathrm{f}}/3}\Big[\frac{1}{\chi n_{\mathrm{d}}}\Big]^{1/3} \,.
\end{equation}

First, we consider an ``extended MRN model" with $\chi = 1$, where nearly all of the dust mass resides in aggregates. Because most of the MRN mass is in $0.25\,\mu\text{m}$ grains, we consider conglomerate growth from only these constituents. Using $\phi_{\text{d}}$ from Table \ref{tab:conditions}, we find $l_{\text{c}} = 5400\,\text{cm}$ for $a = 100\,\mu\text{m}$ deposition sites.

In starless cores, the observation of the $9.7\,\mu\text{m}$ absorption feature attributed to MRN-sized silicate grains implies that not all of the dust participates in coagulation: $\chi \neq 1$ \citep{draine2003review}. Generally, older clouds have higher $\chi$ and larger conglomerates than younger ones \citep{ormel2009dust}. Numerical models from \cite{weidenschilling1994coagulation} find a range of feasible $\chi$, ranging from $10^{-2}$ to $10^{-8}$ for the most evolved and least evolved starless cores. 

Characteristic separations are shown in Table \ref{tab:nucleationspace} for a range of deposition site sizes and $\chi$. The $l_{\text{c}} \propto \chi^{-1/3}$ dependence means that separation of aggregates varies slowly for changing coagulation assumptions. Therefore, the better-constrained $\phi_{\text{d}}$ and $D_{\text{f}}$ lead to reasonable estimates of distance between deposition sites.

Alternatively, non-MRN particles may be delivered from AGB star outflows and other exogenous phenomena. Spacing between these potential deposition sites depends primarily on the starless core's proximity to the source. While nearness to a luminous source would result in temperatures too high for deposition, AGB stars evolve into White Dwarfs within a few $10^{5}\,\text{yr}$ and then would not affect ambient temperature.

\begin{deluxetable}{ccccc}
\label{tab:nucleationspace}
\tablecaption{Summary of deposition site spacing $l_{\text{c}}(a_{\text{c}})$ for noteworthy values of coagulation fraction $\chi$, assuming conglomerates form from the largest MRN constituents. WR1994 references \cite{weidenschilling1994coagulation}.}
\tablewidth{0pt}
\tablehead{
\colhead{Model} & \colhead{$\chi$} & \colhead{$l_{\text{c}}(100\,\mu\text{m})$} & \colhead{$l_{\text{c}}(1000\,\mu\text{m})$}}
\startdata
Extended MRN & $1.0$ & $5.4\times10^{3}\,\text{cm}$ & $2.5\times10^{4}\,\text{cm}$\\
WR1994 High $\chi$ & $10^{-2}$  & $2.5\times10^{4}\,\text{cm}$ & $1.2\times10^{5}\,\text{cm}$ \\
WR1994 Low $\chi$ & $10^{-8}$ & $2.5\times10^{6}\,\text{cm}$ & $1.2\times10^{7}\,\text{cm}$\\
\enddata
\end{deluxetable}

\section{Accretion onto Cloud Products}
\label{section:deposition}

Provided that deposition occurs, we consider the accretion of \htwo{} ice and the remaining free dust onto an isolated cloud product. We discuss growth to centimeter-sized bodies whose dynamics are independent of the turbulence.

\subsection{Mean Free Path and Local Vapor Pressure}

The mean free path $\lambda$ between gas-gas collisions for a given \htwo{} molecule is calculated by

\begin{equation} \label{eq:meanfreepath}
    \lambda_{\mathrm{\htwo}} = (\sqrt{2}\pi (d_{\mathrm{\htwo}}^{2}n_{\mathrm{\htwo}} + d_{\mathrm{He}}^{2}n_{\mathrm{He}}))^{-1}\,,
\end{equation}

\noindent where $d_{\text{\htwo}} = 290\,\text{pm}$ and $d_{\text{He}} = 260\,\text{pm}$ are the effective sizes of \htwo{} and He. In the fiducial core from Table \ref{tab:conditions}, $\lambda_{\text{\htwo}} = 8 \times 10^{7}\,\text{cm}$. For objects with $a \ll \lambda_{\text{\htwo}}$, growth is not diffusion-limited. Hydrogen deposition, instead of forming dendritic ``branches" akin to terrestrial snowflakes \citep{libbrecht2017physical}, would create spherical ``hailstones.".

If solid hydrogen freezes on dust grains, any initial anisotropies, porous regions, and other deformities will be covered by ice. Therefore, in order to continue deposition, either the incident radiation bath must be lower than we estimated in Section \ref{section:nucleation} or the albedo of \htwo{} ice must be sufficiently high. Otherwise, the object's temperature would rise above $2.8\,\text{K}$ as its geometry approaches spherical symmetry.

Applying Equation \ref{eq:meanfreepath} to dust-dust encounters gives $\lambda_{\text{d}} \gtrsim 10^{14}\,\text{cm}$ for the largest MRN grains. We consider only pairwise collisions with dust of equal or greater size, so conglomerates will have even longer $\lambda$. Thus, turbulent gas dynamics dictate the particles' motion. Sub-micron dust may align with ambient magnetic fields as observed in ISM polarization signatures, but bigger grains are unaffected \citep{draine2011book}.

\subsection{Effects of Starless Core Turbulence}

Reynolds Numbers in starless cores are among the highest of any astrophysical flow, so turbulence is difficult to resolve numerically without draconian approximations \citep{draine2011book}. Additionally, effects such as ambipolar diffusion mean that the requirements for ideal magnetohydrodynamics are unfulfilled \citep{vazquez2005lifetimes}. Nonetheless observations show that starless core turbulence empirically fits the subsonic Kolmogorov spectrum \citep{larson1981turbulence}, so we adopt this characterization for our analysis.

Conveniently, the Kolmogorov framework determines the statistics of turbulent feature strength, size, and frequency \citep{weidenschilling1994coagulation}. The largest starless core eddies are of characteristic size $l_{\text{lg}} \sim 0.2\,R_{\text{sc}}$ and have flow velocities $v_{\text{lg}} \sim \beta c_{\text{s}}$ with $\beta \lesssim 1 $. Thus, they have turnover times $t_{\text{lg}} \sim l_{\text{lg}} / (\beta c_{\text{s}})$ dictating their survival. Kinetic energy is $E_{\text{k}} \sim v_{\text{lg}}^{2}$ and cascades to smaller features until dissipative viscous forces become efficient \citep{weidenschilling1994coagulation}. These minimum-size eddies are characterized by

\begin{eqnarray} \label{eq:smallturbulence}
    \nonumber
    l_{\mathrm{sm}} \sim (\eta^{3}/\rho_{\mathrm{g}}^{3}\epsilon)^{1/4}\,, \\
    \nonumber
    v_{\mathrm{sm}} \sim (\eta\epsilon/\rho_{\mathrm{g}})^{1/4}\,,\\
    t_{\mathrm{sm}} \sim (\eta/\rho_{\mathrm{g}}\epsilon)^{1/2}\,,
\end{eqnarray}

\noindent where $\rho_{\text{g}}$ is the gas density, $\eta = 0.1\,\mu\text{Pa}\cdot\text{s}$ is the dynamic viscosity of cold \htwo{} \citep{souers1986hydrogen}, and $\epsilon \sim v_{\text{lg}}^{3} / l_{\text{lg}}$ is the energy dissipation rate \citep{weidenschilling1994coagulation}. Numerical values for fiducial starless core eddies are given in Table \ref{tab:turbulence}. The length scale $l_{\text{sm}}$ must be larger than $\lambda_{\text{\htwo}}$, so the Kolmogorov spectrum provides a minimum eddy size that is slightly above this theoretical limit.

Now, we consider the dynamics of a particle with mass $m_{\text{p}}$, radius $a_{\text{p}}$, and density $\rho_{\text{p}}$, traveling at velocity $v_{\text{p}}$ through an eddy. Its momentum changes via Epstein drag, $|F_{\text{D}}| = (4\pi \rho_{\text{g}}a_{\text{p}}^{2}v_{\text{th}}v_{\text{p}}/3)$, on response timescale \citep{weidenschilling1994coagulation}

\begin{equation} \label{eq:epstein}
    t_{\mathrm{D}} \sim \frac{m_{\mathrm{p}}v_{\mathrm{p}}}{F_{\mathrm{D}}} 
    \sim \frac{a_{\mathrm{p}}\rho_{\mathrm{p}}}{v_{\mathrm{th}}\rho_{\mathrm{g}}}\,.
\end{equation}

The particle crosses an eddy with turnover time $t_{\text{eddy}}$ if $t_{\text{D}} \gtrsim t_{\text{eddy}}$ \citep{weidenschilling1994coagulation}. Micron-sized dust is too small, and conglomerates are too porous to cross even the smallest eddies. However, if deposition proceeds and $\rho_{\text{p}}$ approaches the density $\rho_{\text{ice}}$ of solid \htwo, then the critical object sizes $a_{\text{sm}}$ and $a_{\text{lg}}$ to cross the smallest and largest eddies are

\begin{equation} \label{eq:smalleddycross}
    a_{\text{sm}} \sim \sqrt{\frac{\rho_{\mathrm{g}}v_{\mathrm{th}}^{2}\eta}{\rho_{\mathrm{ice}}^{2}\epsilon}}\, ,
\end{equation}

\begin{equation} \label{eq:largeeddiescross}
    a_{\text{lg}} \sim \frac{l_{\mathrm{lg}} \rho_{\mathrm{g}} v_{\mathrm{th}}}{\rho_{\mathrm{ice}} c_{\mathrm{s}} \beta}\,. 
\end{equation}

\noindent Assuming aggregates $a \gtrsim 100\,\mu\text{m}$ serve as deposition sites, any conglomerates mantled by \htwo{} cross small eddies unimpeded, yet hailstones are somewhat affected by the turbulence until they reach $a_{\text{lg}} \sim 5\,\text{cm}$. Objects larger than $a_{\text{lg}}$ move independently of all features in the Kolmogorov spectrum.

\begin{deluxetable}{ccccc}
\label{tab:turbulence}
\tablecaption{Turbulent eddy conditions in starless cores. Here, $l_{\text{eddy}}$, $v_{\text{eddy}}$, and $t_{\text{eddy}}$ are the length scale, flow velocity, and turnover time of the turbulent features, respectively, and $a_{\text{cross}}$ is the minimum size for a hydrogen hailstone to be unaffected by Epstein drag in the eddy.}
\tablewidth{0pt}
\tablehead{
\colhead{Eddy Size}& \colhead{$l_{\text{eddy}}$} & \colhead{$v_{\text{eddy}}$} & \colhead{$t_{\text{eddy}}$} & \colhead{$a_{\text{cross}}$}
}
\startdata
Largest & $1.5\times10^{16}\,\text{cm}$ & $10^{4}\,\text{cm/s}$ & $5\times10^{4}\,\text{yr}$ & $5\:\text{cm}$ \\
Smallest & $1.8\times10^{9}\,\text{cm}$ & $30 \,\text{cm/s}$ & $3\:\text{yr}$ & $3\,\mu\text{m}$
\enddata
\end{deluxetable}

\subsection{Constraints on Deposition Rates}

An upper bound on the deposition rate can be found from the flux of gaseous \htwo{} molecules. Barring other growth-modifying phenomena, an object of radius $a$ experiences an average collision rate of hydrogen molecules with thermal velocity $v_{\text{th}}$ as

\begin{equation}
    \Gamma
    =
    n_{\mathrm{\htwo}}A_{\mathrm{g}} v_{\mathrm{th}}\,.
    \label{eq:nsigmav}
\end{equation}

\noindent With sticking fraction $s$, we find the volume growth rate

\begin{equation}
    \frac{dV}{dt} 
    =
    \Gamma (m_{\mathrm{\htwo}}/\rho_{\mathrm{ice}}) s\,,
\end{equation}

\noindent where $m_{\text{\htwo}}$ is the mass of one molecule of \htwo.

Then, letting $dV/dt = A_{\text{s}} (da/dt)$, we find the radial growth rate \citep{seligman2020h2}. With appropriate numerical values, we obtain the timescale $t_{\text{dep}}$ for accretion to create spherical hailstones that cross all eddies

\begin{equation}
    t_{\mathrm{dep}} =
    \Big(\frac{2a_{\mathrm{lg}}\rho_{\mathrm{ice}}}{n_{\mathrm{\htwo}}s\sqrt{2kT_{\mathrm{sc}}m_{\mathrm{\htwo}}}}\Big)
\end{equation}

\begin{eqnarray} \label{eq:tformflux}
    \nonumber
    t_{\mathrm{dep}}
    \approx
    3.5\times 10^{5}\,\mathrm{yr}
    \Big[\frac{a_{\mathrm{lg}}}{5\,\mathrm{cm}}\Big]
    \Big[\frac{n_{\mathrm{\htwo}}}{3\times 10^{6}\,\mathrm{cm}^{-3}}\Big]^{-1}\\
    \Big[\frac{s}{0.5}\Big]^{-1}
    \Big[\frac{T_{\mathrm{sc}}}{2.8\,\mathrm{K}}\Big]^{-1/2} \, .
\end{eqnarray}

\noindent After this time, Epstein drag no longer dominates the hailstone dynamics.

In Equation \ref{eq:tformflux}, we have assumed constant collision rate and sticking fraction over the entire time period. However, deposition releases latent heat $L_{\text{dep}} \sim 0.8\,\text{kJ/mol}$ \citep{roder1973survey}, so the hailstone must radiate this energy to maintain its frigid temperature and add more hydrogen ice. Thus, the heat released by the phase transition itself appends another source term to the energy balance in Equation \ref{eq:ISRF} for cloud products that are growing via accretion.

For hailstones where $a \gg \lambda_{\text{cmb}}$, the efficiencies of absorption $\epsilon_{\text{a}}$ and emission $\epsilon_{\text{e}}$ approach unity. If hydrogen ice objects can decouple from the entire turbulent spectrum on the lifetime of a starless core, they must dissipate the latent heat such that their growth is limited by the flux of incident gas particles. Because of the $T^{4}$ dependence in the Stefan-Boltzmann blackbody emission law, the accretion rate is quite sensitive to the ice temperature.

This thermal constraint reiterates the requirement for the starless core to have excellent shielding from electromagnetic radiation and possibly pockets of adiabatic expansion in order to accommodate the latent heat of deposition. As with the dust conglomerate energy balance, the presence of impurity particles may aid the hailstone in efficiently radiating infrared photons.

Thus far, we have neglected the dissipation of gravitational potential energy. While unimportant at the centimeter scale, it provides an upper bound on cloud product size. We equate the latent heat of sublimation for a cloud product to its gravitational binding energy and simplify, finding

\begin{equation} \label{eq:latentplusg}
    R^{2} = \Big(\frac{3}{4\pi}\Big)\Big(\frac{L_{\text{sub}}}{G\rho_{\text{ice}}}\Big)\,.
\end{equation}

From Equation \ref{eq:latentplusg}, we see that gravitational potential energy only restricts the growth of cloud products larger than $4\times10^{8}\,\text{cm}$. This size is much larger than the hailstones that could form via accretion on the characteristic lifetimes of starless cores.

\subsection{Gravitational Focusing of Dust}

Hydrogen deposition dominates for small hailstones, but gravitational focusing of slow-moving dust may meaningfully contribute to mass accretion for large objects. Using conservation of energy and momentum, the impact parameter $b$ of dust with mass $m_{\text{d}}$ attracted to a body of radius $a \gg a_{\text{d}}$ and density $\rho_{\text{ice}}$ is

\begin{equation} \label{eq:impactparamradius}
    b^{2}
    =
    a^{2}
    +
    \frac{\pi Gm_{\mathrm{d}}\rho_{\mathrm{ice}}a^{4}}{3kT_{\mathrm{sc}}}\, .
\end{equation}

\begin{figure}
\plotone{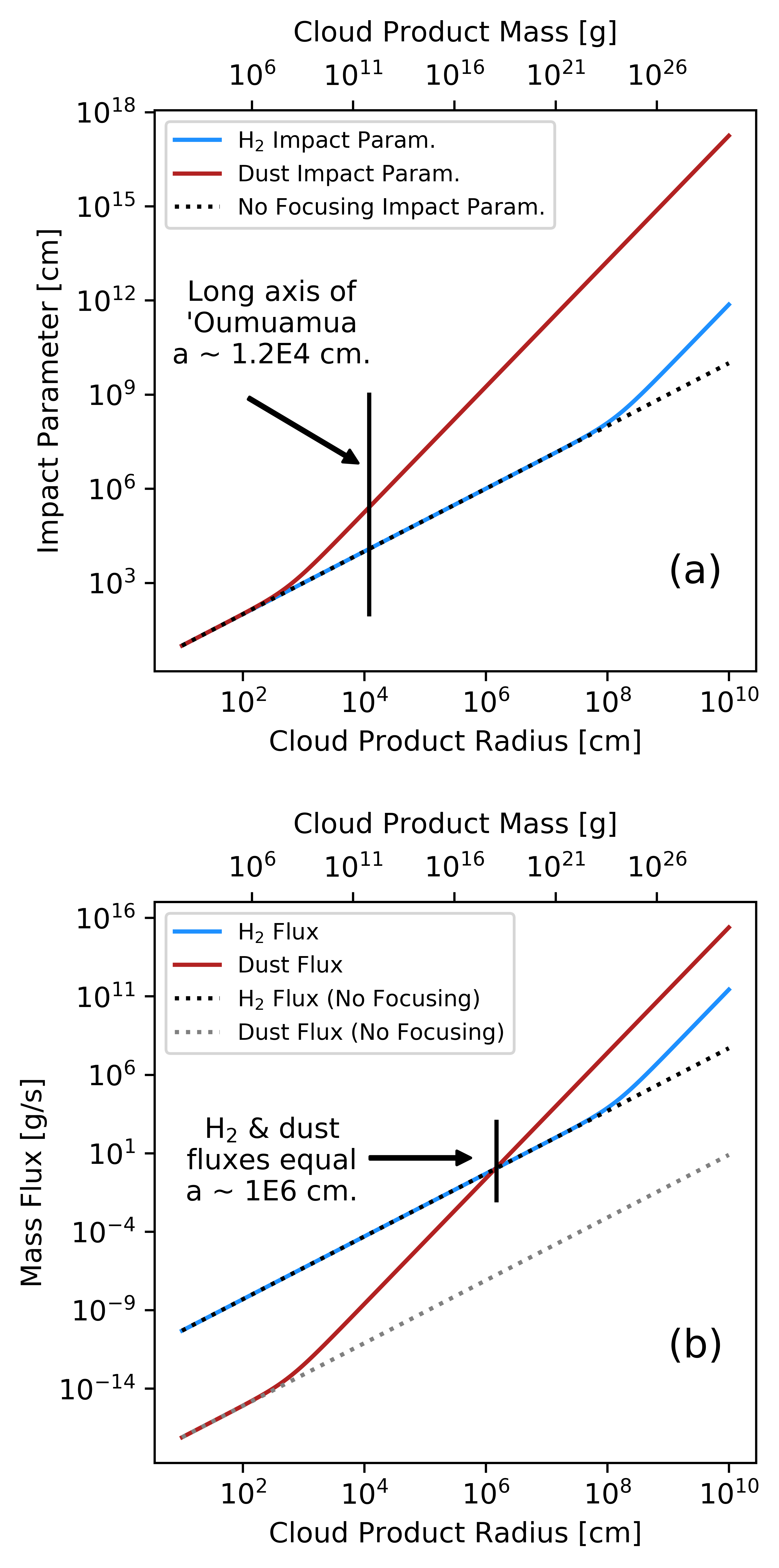}
\caption{Panel (a) shows the impact parameter for a given cloud product radius and mass, assuming $\rho = \rho_{\text{ice}}$. We plot $b$ for both hydrogen gas and MRN dust. Panel (b) shows hailstone mass flux given the impact parameter from Equation \ref{eq:impactparamradius} and the fiducial starless core conditions in Table \ref{tab:conditions}. \label{fig:impactparam}}
\end{figure}

\noindent Here, we have adopted constant $v_{\text{th}}$ at $T_{\text{sc}}$ and assumed low cloud product metallicity via $\rho = \rho_{ice}$. Figure \ref{fig:impactparam} quantifies $b$ and mass flux versus hailstone size. We continue to assume $s_{\text{\htwo}} = \alpha_{\text{\htwo}} = 0.5$, but expect dust $s_{\text{d}} = 1$ from low collision velocities and electrostatic forces.

Next, we incorporate gravitational focusing into the radial growth model. We substitute Equation \ref{eq:impactparamradius} into Equation \ref{eq:nsigmav} for particles of mass $m$, number density $n$, and sticking fraction $s$ colliding with a hailstone of \htwo{} ice to find

\begin{equation} \label{eq:analyticfocusingdiffeq}
    \frac{da}{dt}
    =
   \Bigg(\frac{\sqrt{2}ns}{2\rho_{\mathrm{ice}}}\sqrt{kT_{\mathrm{sc}}m}\Bigg)
    +
    \Bigg(\frac{\sqrt{2m^{3}}\pi Gns}{6 \sqrt{kT_{\mathrm{sc}}}}\Bigg)a^{2}\,.
\end{equation}

\noindent Thus far, we've assumed accretion of only one component, such as either dust or molecular hydrogen. Setting the constant terms in parentheses as $k_{1}$ and $k_{2}$, we get

\begin{equation}
    \frac{da}{dt} = k_{1} + k_{2}a^{2}\,.
\end{equation}

\noindent Physically, the $k_{1}$ term refers to the simple case of accretion onto the geometric cross-section, and the $k_{2}$ term adds the impact of gravitational focusing. Solving with initial condition $a(0) = 0$ since the cloud products begin small, the hailstone size grows as

\begin{equation} \label{eq:analyticfocusingfinal}
    a(t)
    =
    \sqrt{\frac{k_{1}}{k_{2}}}\tan(\sqrt{k_{1}k_{2}} t)\, .
\end{equation}

For two-component accretion of \htwo{} and dust, we let $k_{1} = k_{1, \text{\htwo}} + k_{1, \text{d}}$ and $k_{2} = k_{2, \text{\htwo}} + k_{2, \text{d}}$, where the subscripts correspond to the aforementioned constituents. Because $k_{1, \text{d}} \ll k_{1, \text{\htwo}}$ and $k_{2, \text{d}} \gg k_{2, \text{\htwo}}$, we set $k_{1} \approx k_{1, \text{\htwo}}$ and $k_{2} \approx k_{2, \text{d}}$. We note that $k_{2}$, which represents only dust flux, does not depend on $\rho_{\text{ice}}$.

\begin{figure}
\plotone{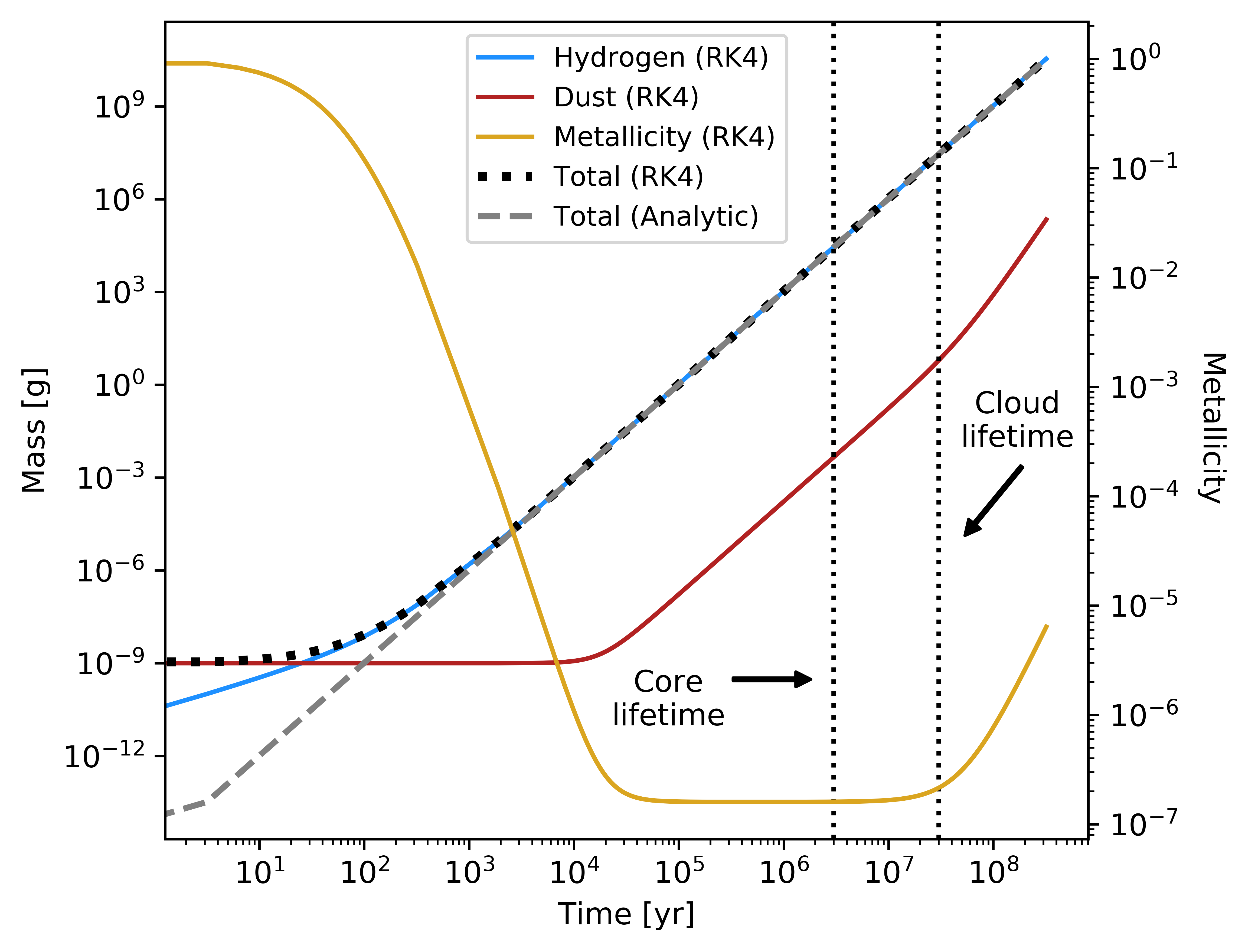}
\caption{Hailstone mass (left-hand axis) versus time for analytic and Runge-Kutta solutions for an isolated body assuming that growth is limited by collisions with ambient gas particles. Cloud product metallicity is shown on the right-hand axis. Hailstones are initialized at $10^{-9}\,\text{g}$ in the numerical scheme, but the analytic solution begins with zero mass. However, the hailstone masses found from the two methods quickly converge for larger cloud product size.}
\label{fig:onebody}
\end{figure}

Figure \ref{fig:onebody} confirms the approximation's validity by comparing Equation \ref{eq:analyticfocusingfinal} to a numerical solution for Equation \ref{eq:impactparamradius}. The latter method considers a single, isolated cloud product and computes the mass flux using the fourth-order Runge-Kutta method. While the analytic solution cannot track hailstone metallicity, the numerical scheme records this information. For both methods, we assume that the accretion rate is limited by the flux of colliding gas particles. Equation \ref{eq:analyticfocusingfinal} eventually leads to runaway growth, but objects do not attain the relevant sizes on starless core lifetimes. 

Gravitational focusing leads to negligible dust accretion at the centimeter-scale, although it is important if cloud products become planetesimal-sized bodies. Nonetheless, linear radial growth for collision-limited accretion of small hailstones means that growth to eddy-crossing objects is nearly independent of initial conglomerate size. However, it also implies that accretion alone cannot explain a decameter-scale cloud product on starless core lifetimes.

\section{Gravitational Attraction and Aggregation of Cloud Products}
\label{section:gravitysnowball}

Here, we examine hailstones that are unimpeded by drag. Namely, pairwise gravitational forces monopolize the dynamics and may permit aggregation to kilometer-sized ``snowballs" \citep{fuglistaler2018solid}. We find different characteristic timescales for accretion of \htwo{} and gravitational aggregation, permitting us to treat these as distinct, separate, and non-overlapping processes.

\subsection{Coagulation Time for Two Objects}

At small size, pairwise impacts between hailstones are determined by an expression with the same form as Equation \ref{eq:nsigmav}, leading to negligible growth in this stochastic manner. For larger objects, however, gravitation increases the collision rate. To fix ideas, we adopt the assumption of constant mass hailstones. Conservation of energy and momentum determine the distance $r_{0}$ from rest over which two objects, each with mass $M$, can gravitationally aggregate in time $\tau$ as

\begin{equation} \label{eq:analyticTwoBody}
    r_{0} = 2\Big(\frac{\sqrt{2GM}\tau}{\pi}\Big)^{2/3}\,.
\end{equation}

Upon contact, resurfacing merges the bodies. However, constant mass is not necessarily true in starless cores because accretion and hailstone-hailstone gravitation may be coincident. To investigate the assumption, we extend the Runge-Kutta model from Section \ref{section:deposition} to encompass two cloud products simultaneously accreting mass according to Equation \ref{eq:analyticfocusingdiffeq} and gravitationally attracting. Objects are initialized at $a_{\text{lg}}$, so we may ignore turbulence.

\begin{figure}
\plotone{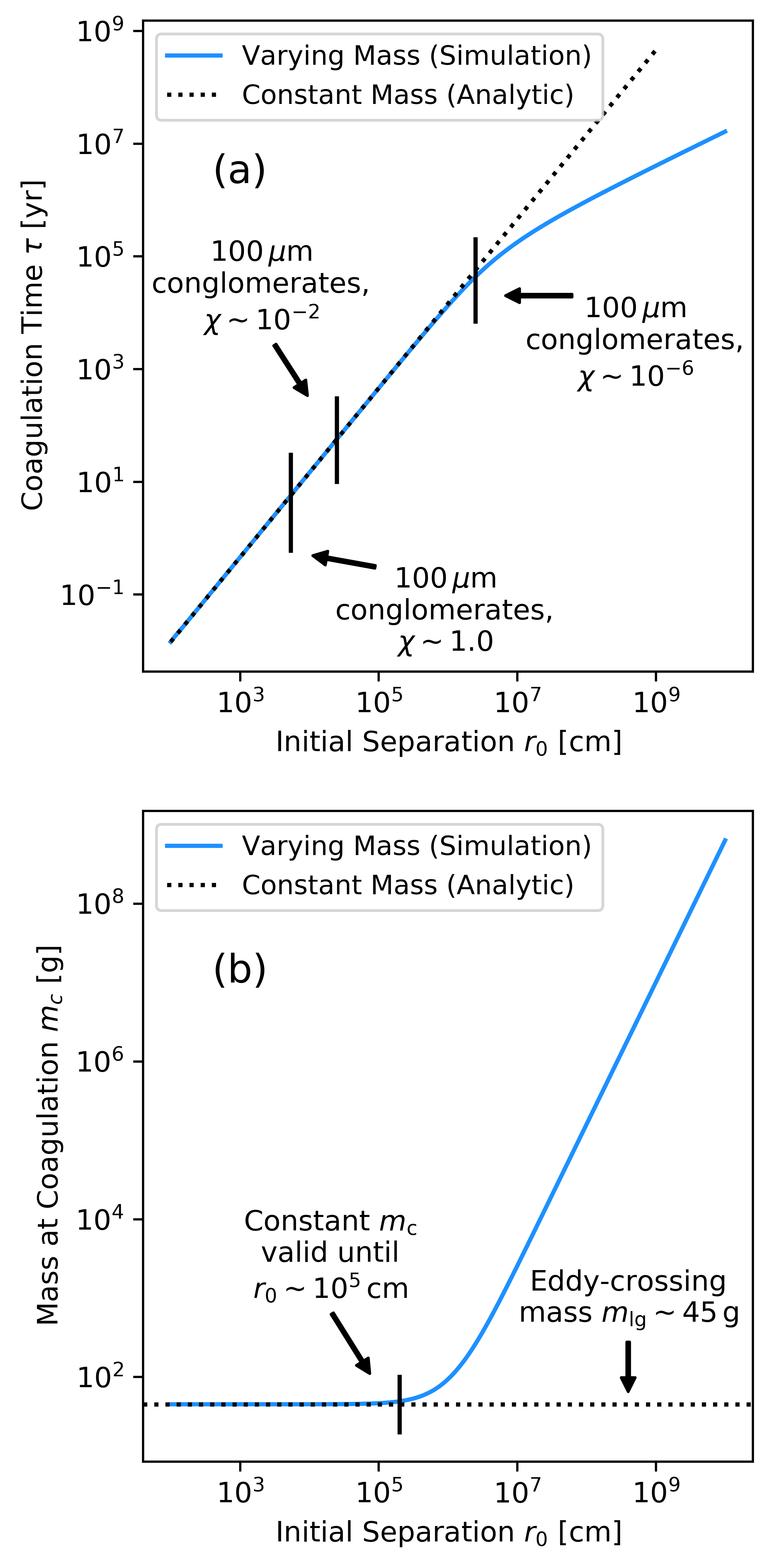}
\caption{Panel (a) shows the coagulation time $\tau$ versus initial separation $r_{0}$, assuming the pairwise gravitation of cloud products with mass $m_{\text{lg}} \sim 45\,\text{g}$. Points corresponding to fiducial deposition site spacing values $\chi$ are marked for reference. Panel (b) shows the mass of objects at coagulation for the same initial separation.
\label{fig:gravitation}}
\end{figure}

Figure \ref{fig:gravitation}(a) shows the coagulation time versus initial separation for eddy-crossing hailstones and the output of Equation \ref{eq:analyticTwoBody} for $M = m_{\text{lg}} = 45\,\text{g}$, the mass of hailstones with radius $a_{\text{lg}}$. For fiducial separations of $100\,\mu\text{m}$ grains in Table \ref{tab:nucleationspace}, the constant mass assumption is fine, but objects with larger $r_{0}$ accrete significant material and coagulate in less time than predicted by Equation \ref{eq:analyticTwoBody}. Similarly, panel (b) shows the mass $m_{\text{c}}$ of the individual bodies at coagulation versus the initial separation $r_{0}$.

Previously, we determined that cloud products must become centimeter-sized before moving independently of the turbulence. Now, we've shown that the coagulation time for two such decoupled objects in starless cores is short enough to treat accretion and pairwise aggregation as distinct processes, consistent with \cite{fuglistaler2018solid}. Thus, Equation \ref{eq:analyticTwoBody} holds for our fiducial separations.

\subsection{Behavior of Multi-Body Systems}

Now, we consider a system of $N$ interacting equal mass bodies. Rearranging Equation \ref{eq:analyticTwoBody}, we find the characteristic attraction timescale $\tau$ for one eddy-crossing object and its nearest neighbor as

\begin{equation} \label{eq:gravityTimescale}
    \tau =
    \Big(\frac{\pi}{4}\Big)
    \Big(\frac{r_{0}^{3/2}}{\sqrt{Gm_{\mathrm{lg}}}}\Big)\, ,
\end{equation}

\noindent For relevant initial separations, mass aggregation by accretion is negligible on timescale $\tau$.

In a system of hailstones with average initial separation $r_{0} = l_{\text{c}}$, the number density is $n_{0} \sim l_{\text{c}}^{-3}$ before coagulation occurs. After one timescale $\tau$, the masses of the remaining bodies are approximately doubled ($m_{1} \rightarrow 2\, m_{\text{lg}}$), and the number density is roughly halved ($n_{1} \rightarrow 0.5\, n_{0}$) if the volume in which they are contained is unchanged. Extending these relationships to find the characteristic timescale $\tau_{i}$ between the $(i-1)^{\text{th}}$ and $i^{\text{th}}$ aggregation, we substitute into Equation \ref{eq:gravityTimescale} and simplify to find

\begin{equation} \label{eq:jeans}
    \tau_{i} =
    \Big(\frac{\pi}{4}\Big)
    (n_{0}Gm_{\mathrm{lg}})^{-1/2}
    =
    \tau\,,
\end{equation}

\noindent which is independent of $i$. Thus, once pairwise gravitation dominates, we expect roughly constant time between each mass-doubling event. Equation \ref{eq:jeans} takes the form of Jeans' collapse: $\tau_{\text{J}} \propto 1/\sqrt{G\rho}$ \citep{jeans1902stability}. Essentially, this result can be understood as the free-fall of a region of hailstones that are unaffected by turbulence. For $l_{\text{c}} = r_{0} = 2.5 \times 10^{4}\,\text{cm}$ in Table \ref{tab:nucleationspace}, we find characteristic $\tau \approx 60\,\text{yr}$. We also note that this result is independent of the ambient starless core temperature.

Next, we solve for the final radius $a_{\text{f}}$ of objects formed after $N_{\text{f}}$ mass doubling events as

\begin{equation}
    a_{\mathrm{f}}
    =
    \Big(\frac{3m_{\mathrm{lg}}2^{N_{\mathrm{f}}}}{4\pi\rho_{\mathrm{ice}}}\Big)^{1/3}\, .
\end{equation}

These aggregated \htwo{} objects, ``snowballs," should have final sizes set by the turbulent spectrum in which the original hailstones are contained. Because small eddies are the most plentiful and the weakest, we expect that features of a few $l_{\text{sm}}$ will determine the most common snowball sizes. Ambient gas in these eddies flows with approximate velocity $1\,\text{m/s}$, so the frontal collisional heating from Epstein drag described in \cite{hoang2020destruction} is unimportant.

Table \ref{tab:aggregation} shows the resulting object radii from coagulating hailstones with size $a_{\text{lg}}$ in eddies of size $l_{\text{eddy}}$, assuming no additional accretion. We find that kilometer-scale cloud products could aggregate within a few  $10^{4}\,\text{yr}$. If the hailstones are attracted to a common center of mass, then the time to coagulate all of the objects in an eddy would be shorter than this estimate. After aggregation, the planetesimal-like objects will collect a few $10^{11}\,\text{g}$ of dust via gravitational focusing on GMC lifetimes. While the newly-aggregated cloud products would still be negligibly metallic, dust will impact the sublimation rate for hydrogen located near these impurities.

\begin{deluxetable}{ccc}
\label{tab:aggregation}
\tablecaption{Estimated snowball radii resulting from the aggregation of all solid \htwo{} hailstones in eddies of size $l_{\text{sm}}$ and $10\,l_{\text{sm}}$.}
\tablewidth{0pt}
\tablehead{
\colhead{Hailstone $l_{\text{c}}$}& \colhead{Eddy Size} & \colhead{Snowball Size $a_{\text{f}}$}
}
\startdata
$5.0\times10^{4}\,\text{cm}$ & $3.0\times10^{9}\,\text{cm}$ & $3.11\,\text{km}$\\
$5.0\times10^{5}\,\text{cm}$ & $3.0\times10^{9}\,\text{cm}$ & $0.31\,\text{km}$\\
$5.0\times10^{6}\,\text{cm}$ & $3.0\times10^{9}\,\text{cm}$ & $0.03\,\text{km}$\\
$5.0\times10^{4}\,\text{cm}$ & $3.0\times10^{10}\,\text{cm}$ & $31.1\,\text{km}$\\
$5.0\times10^{5}\,\text{cm}$ & $3.0\times10^{10}\,\text{cm}$ & $3.11\,\text{km}$\\
$5.0\times10^{6}\,\text{cm}$ & $3.0\times10^{10}\,\text{cm}$ & $0.31\,\text{km}$\\
\enddata
\end{deluxetable}

Finally, we must consider the attraction of bodies to the center of mass in the starless core. For an object of mass $M$ located distance $d$ from the core's center with mass $M_{\text{in}}$ interior to $d$, the Hill Sphere is $r_{\text{H}} \approx d(M/3M_{\text{in}})^{1/3}$. We expect snowballs to form in dark regions with small $M_{\text{in}}$ and thus, $r_{\text{H}} \gg l_{\text{c}}$. Furthermore, the ratio of $r_{\text{H}}$ to the average snowball separation is unchanged for successive mass-doubling events, so coagulation is always unaffected by cloud gravity.

\subsection{Revisiting the Turbulent Spectrum}

So far, we have characterized turbulence via the Kolmogorov framework, despite knowing that the gas is compressible and partially ionized. Instead, one could reasonably choose a sub-Alfv\'enic regime or numerically-simulate the non-ideal magnetohydrodynamics. Nonetheless, only the following general properties of subsonic 3-D flow are important to building hypothetical cloud products:

\begin{enumerate}
    \item Kinetic energy should cascade from eddies with large sizes, high velocities, and long turnover times to weaker, but more plentiful features.
    \item The smallest eddies should be a few $\lambda_{\text{\htwo}}$ in size.
    \item Turbulence should be sufficiently mild so that it neither shatters dust conglomerates nor sublimates \htwo{} ice via heating from Epstein drag.
\end{enumerate}

Since cores are the most quiescent GMC regions and have among the lowest ion fractions in the ISM \citep{draine2011book}, these attributes should apply for any appropriate choice of turbulence. While the Kolmogorov cascade approximately fits observations, an alternate characterization would still find similar plausibility of cloud product formation.

\section{Processing in Interstellar Space}
\label{section:ism}

Now, we examine the feasibility of liberating snowballs from the starless core and discuss processing in interstellar space. We find a lower bound on the size of \htwo{} cloud products that could survive to the Solar System.

\subsection{Release of Snowballs into Interstellar Space}

Once formed, snowballs must leave the starless core to become true interstellar objects. Cloud products in a collapsing, star-forming structure will be destroyed, so only failed cores can produce long-lasting \htwo{} ice. Furthermore, we can eliminate the possibility of a violent ejection because of the energies involved. Assuming snowballs form at the core center, we can find the energy needed for ejection per unit mass $E_{\text{ej}}/M$ as

\begin{equation}
   \frac{E_{\mathrm{ej}}}{M} = G \int_{0}^{\infty} \frac{M_{\mathrm{in}}(r)dr}{r^{2}}\,,
\end{equation}

\noindent where $M_{\text{in}}(r)$ is the core mass interior to radius $r$. Na\"ively assuming a homogeneous sphere, we find $E_{\text{ej}}/M \sim 9.8\,\text{kJ/g}$. For realistic density distributions, energies remain on the same order of magnitude. Because $(E_{\text{ej}}/M) > L_{\text{sub}}$, any ejection event would need to provide multiple times the sublimation energy without destroying the body. 

While gravitational instabilities in starless cores can assume a variety of behaviors and morphologies \citep{fuglistaler2018solid} which could scatter snowballs, the drag heating induced on the tenuous \htwo{} ice makes this event less favored than more gentle dispersal. Moreover, a scattering event may impart a relative velocity to the snowball versus the environment. Assuming the GMC travels with the LSR, an ejected body could have different kinematics compared to \om{} \citep{mamajek2017kinematics, hallatt2020dynamics}.

Due to the barriers imposed by ejections, snowballs must be released by the dissipation of the ambient gas. Re-expansion of cores can occur, especially in unproductive star-forming regions like the Columba and Carina associations \citep{vazquez2005lifetimes}. Denser cores are more apt to collapse, but because their structure is governed by the turbulence from which they originate, a diversity of behavior is expected \citep{draine2011book}.

\subsection{Thermal Sublimation in the Interstellar Medium}

Once the starless core dissipates via radiation pressure \citep{blitz1980gmc}, net sublimation begins. Specifically, \cite{hoang2020destruction} quantify erosion for a pure \htwo{} snowball with temperature $T_{\text{ice}}$ via the crystal lattice oscillation frequency ($\nu_{0} \sim 10^{12}\,\text{s}^{-1}$) as

\begin{equation} \label{eq:sublimation}
    \frac{da}{dt} = -\nu_{0} \Big(\frac{m_{\mathrm{\htwo}}}{\rho_{\mathrm{ice}}}\Big)^{1/3}
    \exp{\Big(\frac{-E_{\mathrm{b}}}{kT_{\mathrm{ice}}}\Big)}\,,
\end{equation}

\noindent where $E_{\text{b}}$ is the binding energy and related to $L_{\text{sub}}$. Evaporative cooling would consistently maintain ice temperature around $3\,\text{K}$ \citep{hoang2020destruction}, so Equation \ref{eq:sublimation} gives the minimum radius of a hydrogen snowball that would survive an ISM journey time of $t_{\text{travel}}$ as

\begin{equation}
    a_{\mathrm{min}} \sim 3\,\mathrm{km} \Big(\frac{t_{\mathrm{travel}}}{30\,\mathrm{Myr}}\Big)\,,
\end{equation}

\noindent in approximate agreement with \cite{hoang2020destruction}. This characteristic timescale independently corroborates the observation of an extreme aspect ratio for \om{} despite its young kinematic age, implying that objects of its class are transient with respect to the Galactic lifetime. Additionally, \cite{hoang2020destruction} discuss enhanced collisional heating on the frontal snowball cross-section due to motion relative to the ambient gas. However, this assumes a high velocity versus the LSR, inconsistent with observations of \om's kinematics.

\subsection{Hydrogen Loss for High Metallicity Remnants}

Hydrogen loss increases snowball metallicity over time because the non-volatile dust grains remain attached to the ISO. If \om{} hosted \htwo{} ice, the surface coverage fraction must have only been around 6\% in the inner Solar System \citep{seligman2020h2}. In contrast, a pure \htwo{} object's non-gravitational acceleration would have exceeded observations by a factor of 16.

Because the binding energy of \htwo{} is higher to impurities than to itself, we expect negligible sublimation from silicate surfaces thinly-covered in \htwo{} \citep{walker2013snowflake}. Thus, Equation \ref{eq:sublimation} is invalid for non-negligible metallicity, even at a temperature above $3\,\text{K}$. Most dust collection occurs when the cloud product is close to its maximum size, so its outer layers will be more metallic. Thus, we could expect shielding of inner \htwo{} and slower mass loss, but it's unlikely to be an order-of-magnitude effect.

Regardless of cloud product metallicity, however, cosmic rays destroy solid \htwo{} and chemically process non-volatile constituents. Thus, for objects dominated by refractory substances, bombardment by energetic particles sets the timescale for hydrogen ice loss. \cite{seligman2020h2} consider an ellipsoidal object, finding that cosmic ray flux with strength similar to the value within the Solar neighborhood can destroy an \om-sized snowball within a few $10^{7}\,\text{yr}$\footnote{Simulations of \om's energy balance can be accessed at \url{https://github.com/DSeligman/Oumuamua_Hydrogen}.}. Thus, cloud products with aggregated silicates will erode rapidly at first from sublimation, then more slowly as cosmic rays become the primary loss process. 

Unless they acquire substantial impurities, kilometer-scale snowballs would vanish within $10\,\text{Myr}$. Thus, any \htwo{} cloud products passing through the Solar System must have been originally larger and formed early enough in GMC lifetime to attract sufficient dust. Consequently, observed interstellar interlopers would be biased towards initially planetesimal-sized bodies.

\subsection{Preservation of a Thin Hydrogen Coating}

Thin hydrogen coatings could survive almost indefinitely on cloud products \citep{walker2013snowflake}. Nonetheless, we must reconcile the preserved \htwo{} mass with the $10^{11}\,\text{g}$ of outgassing material required to explain \om's non-gravitational acceleration \citep{seligman2020h2}. For example, 10\AA{} binding sites \citep{hoang2020destruction} covering a smooth $115\,\text{m}\times115\,\text{m}\times20\,\text{m}$ ellipsoid in a monolayer of molecular hydrogen could only preserve a few grams of \htwo{}. Thus, a cloud product \om{} could not have been this weathered.

Given the strict thermodynamic constraints on forming hydrogen ice cloud products, we evaluate the formation of snowballs directly from \htwo-coated dust. To assemble an ISO with mass $10^{12}\,\text{g}$ directly from MRN particles, approximately $10^{24}$ grains with radius $0.25\,\mu\text{m}$ are required. Only $10^{7}\,\text{g}$ of \htwo{} could be stored on the grains individually, which is insufficient to explain \om's non-gravitational acceleration. While the smallest MRN grains could store sufficient hydrogen, there is no known mechanism to create decameter-sized ISOs from mantled dust grains. Thus, if \om{} were a cloud product, it must have originated with more than a thin \htwo{} coating.

\section{Discussion} \label{sec:discussion}

\subsection{Future Detection of Molecular Hydrogen}

There has yet to be an indisputable discovery of interstellar solid \htwo{}, and most searches have targeted embedded hydrogen within common ISM ices \citep{sandford1993spectroscopic}. While \cite{lin2011interstellar} notes that detection of the H\textsubscript{6}\textsuperscript{+} radical would confirm the presence of UV-bombarded \htwo, the precise spectroscopic signature has not been experimentally-determined. Nonetheless, the future confirmation of processed \htwo{} ice would not necessarily imply the existence of macroscopic ISOs.

To confirm that \om-like objects have an \htwo{} component, in-situ missions would provide the most definitive answers \citep{snodgrass2019interceptor}. Additionally, the forthcoming Vera Rubin Observatory should identify ISOs before their closest approach to Earth and spur follow-up studies of both their intrinsic properties and their dynamics \citep{cook2016realistic}. Narrower density constraints and analysis of a longer trajectory than was available for \om{} could differentiate between radiation pressure, solid hydrogen, and other hypotheses.

\subsection{Solid Hydrogen and Cosmological Evolution}

\begin{figure}
\plotone{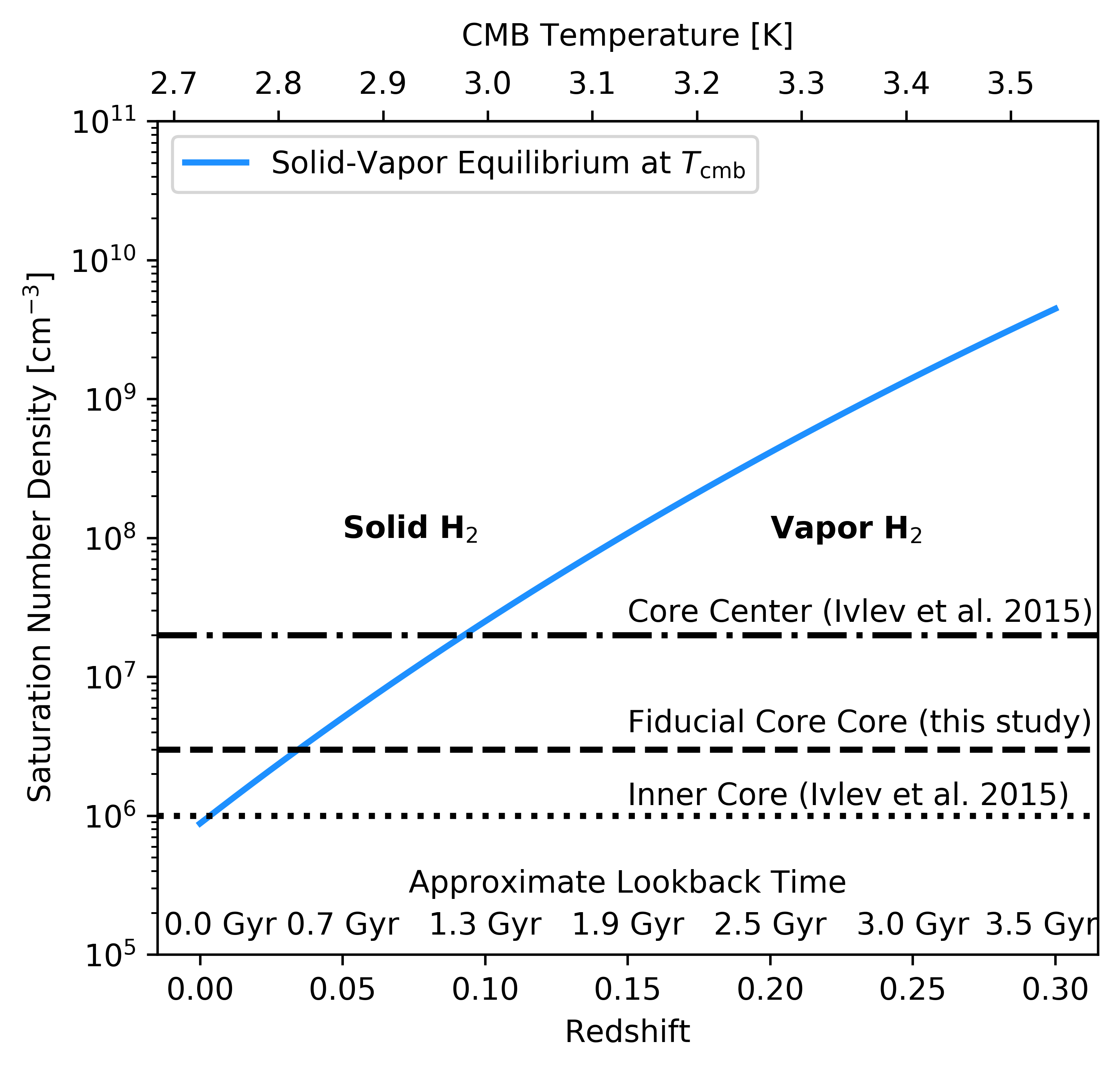}
\caption{Saturation number density of \htwo{} at $T_{\text{cmb}}$, plotted for given redshift and approximate time before present.
\label{fig:redshift}}
\end{figure}

Because CMB temperature scales with cosmological redshift $z$ as $T_{\text{cmb}} = (1 + z)\,2.725\,\text{K} $, solid hydrogen formation is hindered at $z \gtrsim 0.1$ \citep{hoang2020destruction}. Figure \ref{fig:redshift} plots the number density ``ice line" versus redshift, showing that solid \htwo{} could form only within the past $750\,\text{ Myr}$\footnote{Lookback times were found via the Cosmological Calculator at \url{https://home.fnal.gov/~gnedin/cc/} for the default flat Universe (Hubble's constant $68.14\,\text{km/s/Mpc}$, matter density $0.3036$).}. If solid hydrogen ISOs occur in the present epoch, we would expect increasing production in the Universe's far future \citep{white1996dark}.

\subsection{GMC Constraints from Solid Hydrogen ISOs}

If \om-like objects contain dynamically-important \htwo, it would illuminate previously-unobserved conditions in the coldest starless cores. Furthermore, we could constrain the efficiency of the formation process. \cite{do2018interstellar} estimates the ISO number density as $n_{\text{iso}} \sim 0.2\,\text{AU}^{-3}$ with 3.5 years of Pan-STARRS data, so we update this to $n_{\text{iso}} \sim 0.1\,\text{AU}^{-3}$ to account for additional time without another \om-like object. We consider the mass of snowballs $m_{\text{iso}}$ at their maximum, before processing in the ISM. Due to \om's low velocity relative to the LSR, we assume that the Earth is passing through a group of cloud products that have barely moved from that fiducial Galactic orbit since their formation. Therefore, the efficiency $e$ is

\begin{equation}
    e \approx 
    \frac{n_{\mathrm{iso}} m_{\mathrm{iso}}}{\rho_{\mathrm{cl}}}\,,
\end{equation}

\noindent where $\rho_{\text{cl}}$ is the GMC mass density. Thus, we find

\begin{eqnarray}
    e \sim 2.5\times10^{-5}
    \Big[\frac{n_{\mathrm{iso}}}{0.1\,\mathrm{AU}^{-3}}\Big]
    \Big[\frac{r_{\mathrm{iso}}}{2\,\mathrm{km}}\Big]^{3}
    \Big[\frac{10^{3}\,\mathrm{cm}^{-3}}{n_{\mathrm{\htwo}}}\Big]\,.
\end{eqnarray}

Thus, ISO formation in a snowball-spawning cloud would be approximately 400 times less efficient than star formation \citep{blitz1980gmc}. If \om-like objects have solid \htwo{}, population-level statistics will considerably refine our estimate.

\subsection{Catalytic Processes as an Alternative Mechanism}

In Section \ref{section:deposition}, we considered formation where freezing \htwo{} attaches mostly onto other \htwo{} molecules. However, another hypothetical ISO assembly mechanism may involve a catalytic process where deposition only occurs on dust grains, perhaps at slightly higher temperature than $T_{\text{gr}} = 3\,\text{K}$. If solid hydrogen particles are formed and released, then the cross-section $A_{\text{g}}$ in Equation \ref{eq:nsigmav} is constant. In time $t$, one grain of radius $a$ can catalyze deposition for mass

\begin{equation} \label{eq:catalysis1}
    M_{\mathrm{cat}} = 
    n_{\mathrm{\htwo}}\pi a^{2}v_{\mathrm{th}}m_{\mathrm{\htwo}}st\,.
\end{equation}

\noindent Assuming the MRN distribution of dust sizes ($dn/da \propto a^{-3.5}$), nearly all of the surface area will be contained in the smallest grains. Therefore, we consider only these grains as catalysis sites in our order-of-magnitude computation. Thus, we can determine the GMC volume with characteristic side length $l$ required to produce the amount of \htwo{} required for \om's survival in the ISM.

\begin{equation}\label{eq:catalysis2}
    l = \Big[\frac{n_{\mathrm{\htwo}}\pi a_{\mathrm{d}}^{2}v_{\mathrm{\htwo}}m_{\mathrm{\htwo}}stn_{\mathrm{d}}}{M}\Big]^{1/3}
\end{equation}

Table \ref{tab:catalysis} shows a few example box sizes required for the production of $10^{15}\,\text{g}$ ISOs. Coincidentally, these results are close to the \htwo{} mean free path that creates natural length scales. Despite this correspondence, there is no known chemical process that would both freeze and release \htwo{}. Furthermore, free hydrogen crystals would experience the ambient photon bath discussed Section \ref{section:nucleation} and rapidly sublimate. Thus, catalytic processes, if they occur at all, probably cannot produce the multi-kilometer objects required for survival to the Solar System.

\begin{deluxetable}{cccc}
\label{tab:catalysis}
\tablecaption{Summary of relevant volumes and timescales to form an ISO with $10^{15}\text{g}$ of \htwo{} mass via an unknown catalytic process.}
\tablewidth{0pt}
\tablehead{
\colhead{Dust Size}& \colhead{Formation Time} & \colhead{Box Side $l$} & \colhead{Box Side $l/l_{sm}$}}
\startdata
0.005 $\mu\text{m}$ & $3\times10^{5}$ yr & $5.0\times10^{9}\,\text{cm}$ & 2.8 \\
0.25 $\mu\text{m}$ & $3\times10^{5}$ yr  & $9.5\times10^{9}\,\text{cm}$ & 5.3  \\
0.005 $\mu\text{m}$ & $3\times10^{6}$ yr & $2.3\times10^{9}\,\text{cm}$ & 1.3 \\
0.25 $\mu\text{m}$ & $3\times10^{6}$ yr  & $4.5\times10^{9}\,\text{cm}$ & 2.5 \\
\enddata
\end{deluxetable}

\subsection{Implications for the Population of ISOs}

We wish not to speculate on the broader population of objects from just \om, but rather hypothesize generic attributes based on the characteristics of our outlined formation mechanism. Because the approximate lifetime of the solid \htwo{} component is $50\,\text{Myr}$, cloud products of older age would be left solely with refractory substances. The remnant population's number density depends primarily on the exact lookback time at which the formation of their solid hydrogen predecessors became feasible. 

Additionally, ISOs with active \htwo{} sublimation would be concentrated near GMCs with failed cores. Thus, the spatial distribution of young \htwo{} snowballs would be anisotropic, whereas the eroded remnants would form a homogeneous Galactic background. Both new and old cloud products should pass through the Solar System, but only the former should accelerate non-gravitationally. Within the next decade, a larger sample of objects will further constrain the population-level statistics of ISOs.

\section{Summary and Conclusions} \label{section:conclusion}

\begin{deluxetable*}{cp{2.5cm}p{4cm}cp{4cm}}
\label{tab:summary}
\tablecaption{Summary of size regimes considered in the formation of solid \htwo{} cloud products. For each characteristic object, we state the physical processes, relevant formation timescales, and potential barriers to formation.}
\tablewidth{0pt}
\tablehead{
\colhead{Size Regime} & \colhead{Object Type} & \colhead{Relevant Processes} & \colhead{Timescale} & \colhead{Possible Barriers}
}
\startdata
\\
$0.005\,\mu\text{m} < a < 0.25\,\mu\text{m}$ & \centering MRN-Sized ISM Dust Grains & \centering GMC complex and starless core formation. & $3\,\text{Myr}$ & {\centering None. \\}\\
\\
\hline \\
$0.25\,\mu\text{m} < a < 100\,\mu\text{m}$ & \centering Large Dust Conglomerates & \centering Grain coagulation and possible exogenous delivery. & $1\,\text{Myr}$ & {\centering Starless cores must support formation of large grains.\\}\\ 
\hline \\
$100\,\mu\text{m} < a < 5\,\text{cm}$ & \centering Solid Hydrogen Hailstones & \centering Deposition limited by hydrogen gas flux. & $350\,\text{kyr}$ & {\centering $T_{\text{sc}}$ must be below $2.8\,\text{K}$. Latent heat must dissipate.\\} \\
\hline \\
$5\,\text{cm} < a < 5\,\text{km}$ & \centering Aggregating Snowballs & \centering Gravitational coagulation of eddy-crossing hailstones. & $10\,\text{kyr}$ & {\centering Initial separation must be small for gravitation.\\} \\
\hline \\
$100\,\text{m} < a < 5\,\text{km}$ & \centering Interstellar Snowballs & \centering Thermal sublimation and cosmic ray bombardment. & $50\,\text{Myr}$ & {\centering ISO must aggregate enough dust to preserve \htwo.\\} \\
\hline \\
$a \sim 100\,\text{m}$ & \centering \om-Like Interlopers & \centering Non-gravitational acceleration, outgassing. & $5\,\text{yr}$ & {\centering None.\\}\\
\\
\enddata
\end{deluxetable*}

Beginning from the observation that \htwo{} outgassing would explain many of \om's exceptional attributes \citep{seligman2020h2}, we analyze the feasibility of a generic ISO formation mechanism in GMCs. Because of the severe thermodynamic criteria for deposition, only failed cores in unproductive star-forming regions could possibly spawn these bodies.

Table \ref{tab:summary} summarizes the size regimes and physical processes we have considered in the hypothetical production of generic \om-like ISOs. Specifically, we examine starless cores for the possibility that they may spawn solid \htwo{} cloud products. We find that the strongest barrier is the temperature requirement for \htwo{} deposition. Because molecular clouds are not opaque in the infrared, both the CMB and radiation from warm dust can penetrate to all relevant visual extinctions. Nonetheless, it is possible that deposition could occur on non-ideal grain conglomerates.

If deposition occurs, hailstones that cross all turbulent eddies in the Kolmogorov spectrum could form within $350,000\,\text{yr}$ given efficient dissipation of the latent heat. At this macroscopic size, pairwise attraction between cloud products becomes the catalyst for mass aggregation. In the ISM, thermal sublimation determines the hydrogen component's lifetime and introduces a minimum radius for survival given a travel time. The \htwo{} fraction at the Solar encounter is determined by the interstellar journey time and the mass of silicates attracted via gravitational focusing. Young objects typically have inbound velocities near the Local Standard of Rest, as they have not experienced strong scattering events \citep{hallatt2020dynamics}.

Currently, solid hydrogen is the only hypothesis that explains all of \om's observed properties -- non-gravitational acceleration, lack of cometary volatiles, high aspect ratio, inbound kinematics, and number density -- as generic to a class of naturally-forming objects. Thus, despite its exotic nature, the \htwo{} cloud product remains as compelling as any other interpretation. The composition of \om-like ISOs will be testable with forthcoming observational surveys, and the answer may illuminate a previously-unknown avenue of small body formation.

\vspace{5mm}

\noindent We thank Daniel Pfenniger for a detailed peer-review, which greatly improved the manuscript. Additionally, we thank Bruce Draine and Dave Hollenbach for the engaging conversations on ISM physics and astrophysical dust. We also thank Joel Ong for the comment on the figure caption corresponding to the accretion model. Finally, we are grateful to Darryl Seligman, Mark Walker, Fred Adams, and Christopher Lindsay for providing comments from closely-reading our draft paper. This research used the \texttt{numpy} \citep{harris2020array} and \texttt{matplotlib} \citep{hunter2007matplotlib} packages in \texttt{python}.

\bibliography{bibliography}{}
\bibliographystyle{aasjournal}

\end{document}